\documentclass[12pt, preprint]{aastex}
\usepackage{natbib}
\usepackage{amsmath}
\usepackage{rotating}

\def\apj{ApJ}

\def\apjl{ApJ}
\def\aap{A\&A}
\def\nat{Nat}
\def\mnras{MNRAS}
\def\ssr{SSR}
\def\planss{Planet. Space. Sci.}
\def\icarus{Icarus}

\def\me{$M_{\oplus}$}

\begin{document} 

\title{The Heavy Element Masses of Extrasolar Giant Planets, Revealed} 
\author{Neil Miller and Jonathan J. Fortney\altaffilmark{1}}
\affil{Department of Astronomy and Astrophysics, University of California, Santa Cruz}
\email{neil@astro.ucsc.edu}
\altaffiltext{1}{Alfred P. Sloan Research Fellow}

\begin{abstract}
We investigate a population of transiting planets that receive relatively modest stellar insolation, indicating equilibrium temperatures $< 1000$ K, and 
for which the heating mechanism that inflates hot Jupiters does not appear to be significantly active.
We use structural evolution models to infer the amount of heavy elements within each of these planets.  
There is a correlation between the stellar metallicity and the mass of heavy 
elements in its transiting planet(s).  It appears that all giant planets posses a minimum of $\sim$ 10-15 
Earth masses of heavy elements, with planets around metal-rich stars having larger heavy element
masses.  There is also an inverse relationship between the mass of the planet and the metal enrichment 
($Z_{\textrm pl} / Z_{\textrm star}$), which appears to have little dependency on the metallicity of the star.  Saturn- and Jupiter-like enrichments above solar composition are a hallmark of all the gas giants in the sample, even planets of several Jupiter masses.
These relationships provides an important constraint on planet formation, and suggests large amounts of heavy elements within planetary H/He envelopes.  
We suggest that the observed correlation can soon also be applied to inflated planets, such that the interior heavy element abundance of these planets could be estimated, yielding better constraints no their interior energy sources.  
We point to future directions for planetary population synthesis models and suggest future correlations.  This appears to be the first evidence that extrasolar giant planets, as a class, are enhanced in heavy elements.
\end{abstract}
\maketitle 


\section{Introduction}
Transiting exoplanets are valuable for planetary characterization because they allow us to measure their masses through stellar
radial velocity or other dynamical measurements, as well as their radii from the transit light curve.  Together, these yield a planet's bulk density.  In principle, this information could be used to determine a planet's
composition as increasing the mass fraction of heavy elements increases the density.  This apparently straightforward method has been difficult to implement, however.
Transit observations have revealed that most of the highly irradiated ``hot Jupiters'' 
are inflated to large radii beyond what is expected from simple models.  
The reason for this effect has not been determined;
a variety of additional internal energy sources or contraction-stalling mechanisms have been proposed 
\citep{Guillot02, Jackson09, Batygin2011, Chabrier07c, Arras2010}.  
Since an inflated radius decreases a planet's density, the heating mechanism acts in opposition to the effect of adding heavy elements to the planet.  Therefore interior composition for a transiting planet is generally left entirely unknown, unless planets are found to be dramatically overdense \citep[e.g.][]{Sato05,Fortney06,Leconte2009}.

For the over-inflated planets it is possible to find a relation between the heavy elements in the planets and the metallicity
of the stars by making an {\it assumption} about the relationship between the incident stellar radiation and the unknown power input into the planet \citep{Guillot06}.  
A similar relationship has been found by using \emph{ad hoc} enhanced atmospheric opacity to slow planetary contraction \citep{Burrows07}.
These studies are intriguing, although the resulting planet-star metallicity relationship is dependent on the assumed behavior of the unknown radius inflation mechanism.  Since it is well-known that our solar system's four giant planets possess at least $10-15$~$M_{\oplus}$ of heavy elements within their interiors, making them enriched compared to the Sun's composition \citep{FortneyNettelmann10}, it is paramount to determine the composition of giant exoplanets to understand the structure and formation of these planets as a class of astrophysical objects.

Empirically, the unknown heating mechanism affects the close-in planets at high incident stellar flux \citep{Kovacs10} or planet $T_{\textrm{eff}}$ \citep{Laughlin2011}.  
This is shown in Figure 1, where we plot the observed planet
radii as a function of their average incident stellar flux.  A thermal evolution model for a 1 Jupiter-mass (1 $M_{\rm J}$) planet with no
extra heating source other than the effect of incident radiation on the planet's atmosphere
is plotted with a heavy element core (25 Earth masses, $M_{\oplus}$, dotted) and without (solid) \citep{Miller09}.  At approximately $\langle F \rangle = 2 \times 10^8$ erg s$^{-1}$ cm$^{-2}$, the sample is divided into
two regions \footnote{$\langle F \rangle = 2 \times 10^8$  erg s$^{-1}$ cm$^{-2}$ corresponds to an equilibrium temperature of 990 K for an Bond albedo of 0.1 and efficient heat transport between the day and night side.  This temperature is quite similar to that for which Ohmic heating is suggested to become important \citep{Batygin2011}}: those in the higher flux region, most of which require extra heating to explain their radii, and those in the lower flux region, in which no inflated planets are found.
This empirically suggests that the heating mechanism does not significantly
contribute to the energy budget at low incident flux.
Therefore, for this sample of 14 transiting exoplanets, we can neglect the heating mechanism and 
use our structural evolution models to estimate a planet's composition.

\begin{figure}[!t]
\begin{center}
\includegraphics[width=0.98\columnwidth]{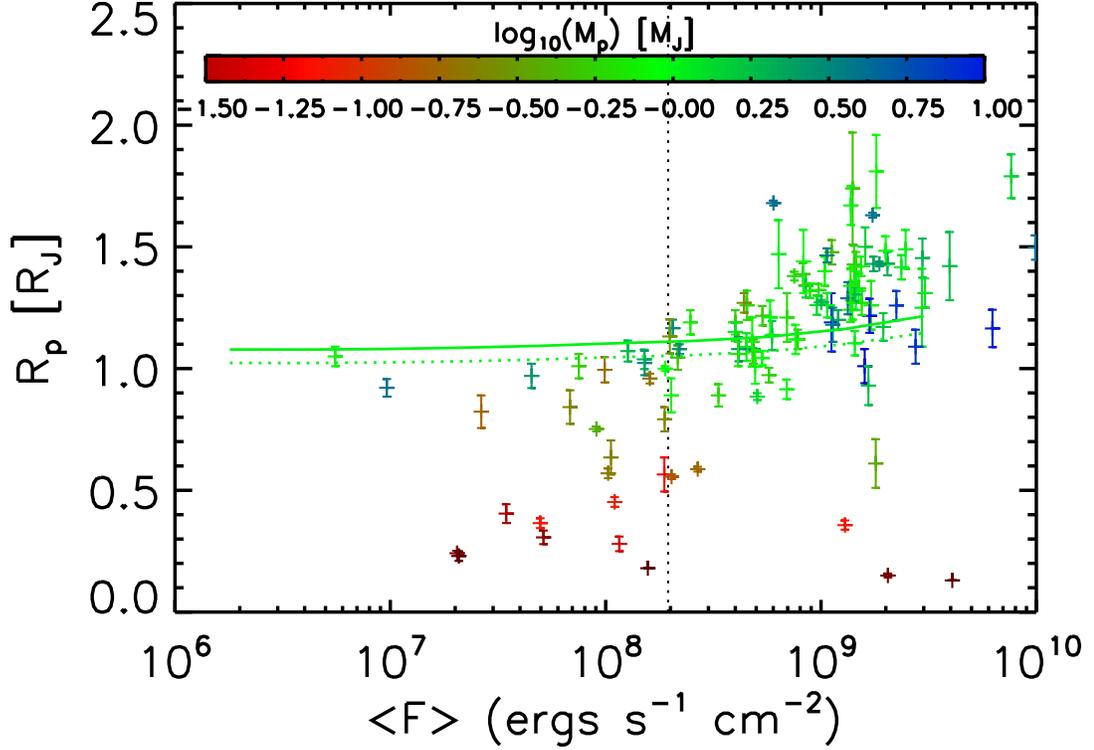}
\caption{
Planet radius as a function of average incident stellar flux.  
Planets are colored according to their mass.
Model planet radii are plotted for a 1 $M_{\rm J}$ planet at 4.5 Gyr
without a core (solid) and with a 25 $M_{\oplus}$ core (dotted) \citep{FortneyMarleyBarnes07,Miller09}.
Although the extra heating source is not well-determined, it is clear that it is more
important at larger incident fluxes.  
We choose a cutoff of $\langle F\rangle < 2 \times 10^8$ erg s$^{-1}$ cm$^{-2}$ in order to obtain the largest 
sample of non-inflated planets.  This corresponds to a planetary $T_{\rm eq}\lesssim 1000$ K.}
\end{center}
\end{figure}

\begin{figure}[!t]
\begin{center}
\includegraphics[width=0.98\columnwidth]{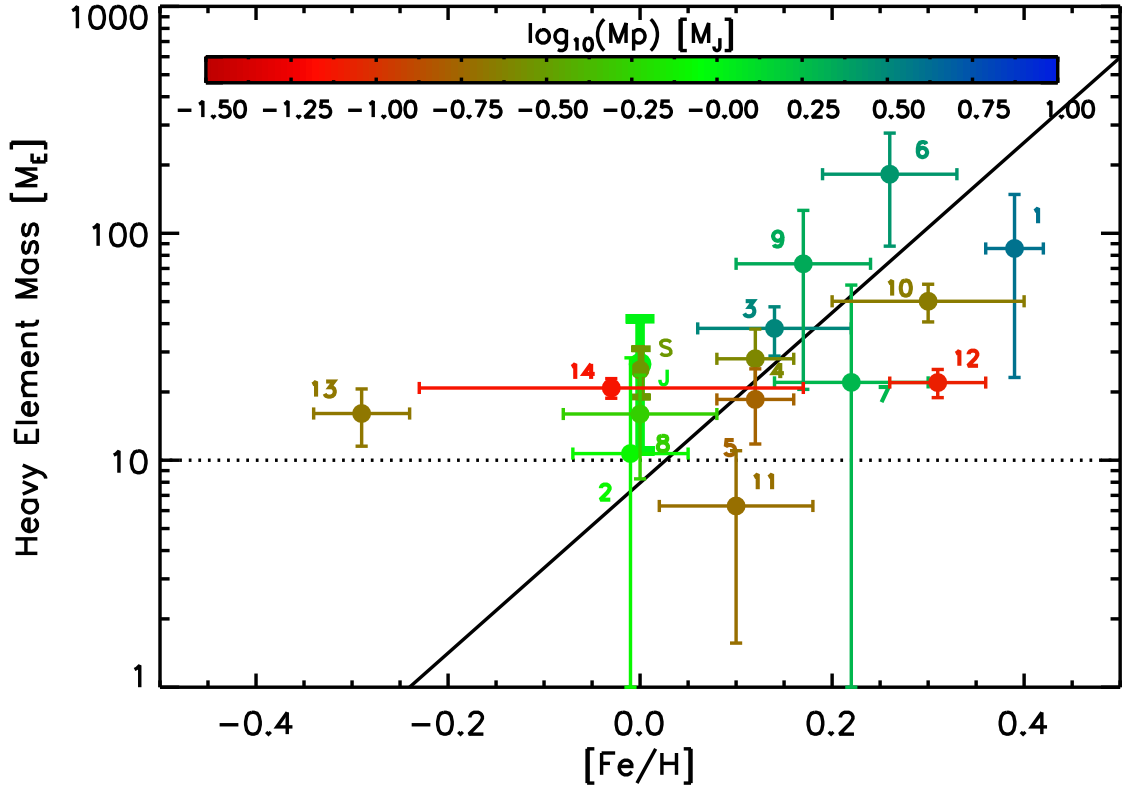}
\caption{
The stellar metallicity and inferred planet heavy element mass for exoplanets within our incident flux cut.  
The required heavy elements are from the ``Average Case" in Table 1. (See text.)
Planets are numbered corresponding to the entries in Table 1.  
The rarity of gas giants around metal-poor stars is well established \citep{Fischer05}.  
Using a least-squares fit, we find the relation $\log M_Z = (0.82 \pm 0.08) + (3.40 \pm 0.39) [\textrm{Fe/H}]$ and a reduced Chi-squared value of 1.95.  The fit excludes HAT-P-12b (planet 13) and includes Jupiter and Saturn.  However, we do not expect this relation to hold 
at the lowest metallicities, where it may become flat at $\sim$~10-15 $M_{\oplus}$.
}
\end{center}
\end{figure}

\section{Model and Method}
We consider two limiting types of planetary structures.  
We consider planets where all of the heavy elements are in an inert core with an adiabatic solar metallicity convective envelope above (layered model).  
We also consider a structure where heavy elements are uniformly mixed with the hydrogen and helium and the planet is fully convective (mixed model).  
The primary heavy element composition is a mixture of 50\% rock and 50\% ice using the equation of state ANEOS \citep{ANEOS}.  
By considering the two extreme cases of having all of the heavy element masses in the core or envelope, we bracket possible interior models of giant planets.  For Jupiter, models that match gravity field constraints generally find that most of its heavy elements are in the envelope while for Saturn most are in the core \citep{FortneyNettelmann10}.

A complete description of the thermal evolution model can be found in \citet{FortneyMarleyBarnes07} and \citet{Miller09}.  Briefly, planets are composed of up to three components: 1) an inert core, 2) an adiabatic convective envelope (where heavy elements may be mixed in), and 3) a solar-metallicity non-grey atmosphere model
\citep{FortneyMarleyBarnes07} that includes the atmospheric extension to the transit radius.  
The primary effect of heavy elements either in the core or in the convective envelope is mainly to decrease the planet's radius at every time.  

For each planet, the amount of heavy elements is determined under the constraint that the predicted model transit radius agrees with the observed
radius at the observed age and incident flux.  
The average incident flux that a planet receives is given by
\begin{equation}
\langle F \rangle = \frac{L_*}{4 \pi a^2 \sqrt{1-e^2}}
\end{equation}
where $L_*$ is the luminosity of the star, $a$ is the semi-major axis
of the orbit and $e$ is the eccentricity of the orbit.
This analysis was performed on all planets that met our 
average incident flux cut $\langle F \rangle < 2 \times 10^8$ erg s$^{-2}$ cm$^{-2}$ and had a mass greater than 20 $M_{\oplus}$--since our model is primarily designed to describe giants with masses greater than Neptune.

Note these heavy element masses should be taken as {\it minimum masses} since if the planet is internally heated or if higher atmospheric opacities (due to metal-enhanced atmospheres) slow the cooling \citep{Ikoma06,Burrows07}, then a planet would have more heavy elements than found here.

The required heavy element mass to fit the radius is determined as the average of the layered and mixed cases.  
Each of the the observed system parameters ($R_p$, age, $a$, $M_p$) has an associated error on its published value.  The propagated error on the heavy element mass ($\sigma_H$)
is given by:
\begin{eqnarray}
\sigma_H^2 &=& \left| \frac{\partial M_c}{\partial R_p} \right|^2 \sigma_{R_p}^2 + \left| \frac{\partial M_c}{\partial \textrm{Age}} \right|^2 \sigma_{\textrm{Age}}^2 
           + \left| \frac{\partial M_c}{\partial a} \right|^2 \sigma_a^2 \nonumber \\&&+ \left| \frac{\partial M_c}{\partial M_p} \right|^2 \sigma_{M_p}^2 + \left(\frac{M_c - M_{\textrm{env}}}{2}\right)^2 
\end{eqnarray}
where $\sigma_{R_p}$, $\sigma_{\textrm{Age}}$, $\sigma_a$, and $\sigma_{M_p}$ are the observationally
determined errors in planet radius, system age, semi-major axis, and planet mass respectively.  
The derivatives $\frac{\partial M_c}{\partial X}$ (calculated at the observed planet parameters assuming core heavy elements) describe the sensitivity of the predicted heavy element mass with respect to changes in a given parameter, $X$.  The final term of the expression is the uncertainty due to the unknown structure of the planet.  $M_c$ and $M_{\textrm{env}}$ are the predicted heavy element masses if the heavy elements are within the core, or the envelope, respectively.

We use the metallicity of the star [Fe/H] as given in each paper in Table 1.  For each system, we compute the heavy element mass fraction $Z_{\textrm{star}} \equiv 0.0142 \times 10^{[Fe/H]}$ - assuming that the total heavy element composition of other systems
scales with their iron abundance, normalized to the solar metalicity as in \citet{Asplund09}.  

\section{Findings}
In Figure 2 we plot the stellar metallicity, [Fe/H], against the planet heavy element mass for each of these systems.  
Using a least squares fit, we find that  $\log M_Z = (0.82 \pm 0.08) + (3.40 \pm 0.39) [\textrm{Fe/H}]$ for stars with $\textrm{ [Fe/H]} > -0.05$. 
The reduced Chi-squared value of 1.95 implies that not all of the scatter can be explained by observational error.  We expect a fairly flat relation (the dotted line in Figure 2) at subsolar stellar metallicity if 10-15 \me\ of heavy elements are needed to trigger planet formation.
In Table 1 we list the planets and observed parameters used.  
For each planet, we list the average predicted heavy elements between the core model and mixed model with the 50-50 rock-ice composition.  

We have examined the sensitivity of our findings to alternate choices for the heavy element EOS, and the differences are small.  
On the low density EOS end, we have used 100\% water \citep{ANEOS}, and on the high density end 2/3 rock \citep{ANEOS}  and 1/3 iron \citep{SESAME}.  
These generally lead to  $Z_{\textrm{pl}}/Z_{\textrm{star}}$ that are 10-20\% larger, for the water EOS, and 10-20\% smaller, for the rock/iron EOS,
 than those found in the last column of Table 1.  
For example, for the HAT-P-17b average case, the best-fit heavy element mass of $16.0\pm7.7$ \me\ increases to $17.7\pm4.8$ \me\ for pure water, 
and $14.1\pm3.9$ \me\ for rock/iron.  This yields $Z_{\textrm{pl}}/Z_{\textrm{star}}$ values of $7.4\pm2.5$ and $5.9\pm2.0$ for the water and rock/iron cases, respectively, very similar to the $6.7\pm3.5$ value in Table 1.

Perhaps the clearest way of looking at this sample is to compare the planet mass against the inferred heavy element mass or heavy element enrichment ($Z_{\rm pl} / Z_{\textrm star}$), as shown in Figure 3.   
The mass of heavy elements appears to increase with planet mass.  On the other hand, the heavy element \emph{enrichment}  decreases with increasing planet mass consistent with the pattern found in the solar system's four giant planets \citep{FortneyNettelmann10}.  
All of these planets are consistent with being enriched in heavy elements and many of these must be enriched significantly.

Within the cluster of Saturn-like planets around 0.2 $M_J$, the most heavy element-rich planet (10, CoRoT-8b) orbits the most metal-rich parent star.  For the planets more massive than Jupiter, the planets harbor large amounts of heavy elements, and orbit around metal-rich stars, which explains some of the steep slope from Figure 2. 
It is interesting to note that the relationship of $Z_{\textrm{pl}}/Z_{\textrm{star}}$ as a function of planet mass, appears fairly independent of stellar [Fe/H].

\begin{figure}[!b]
\begin{center}
\includegraphics[width=0.98\columnwidth]{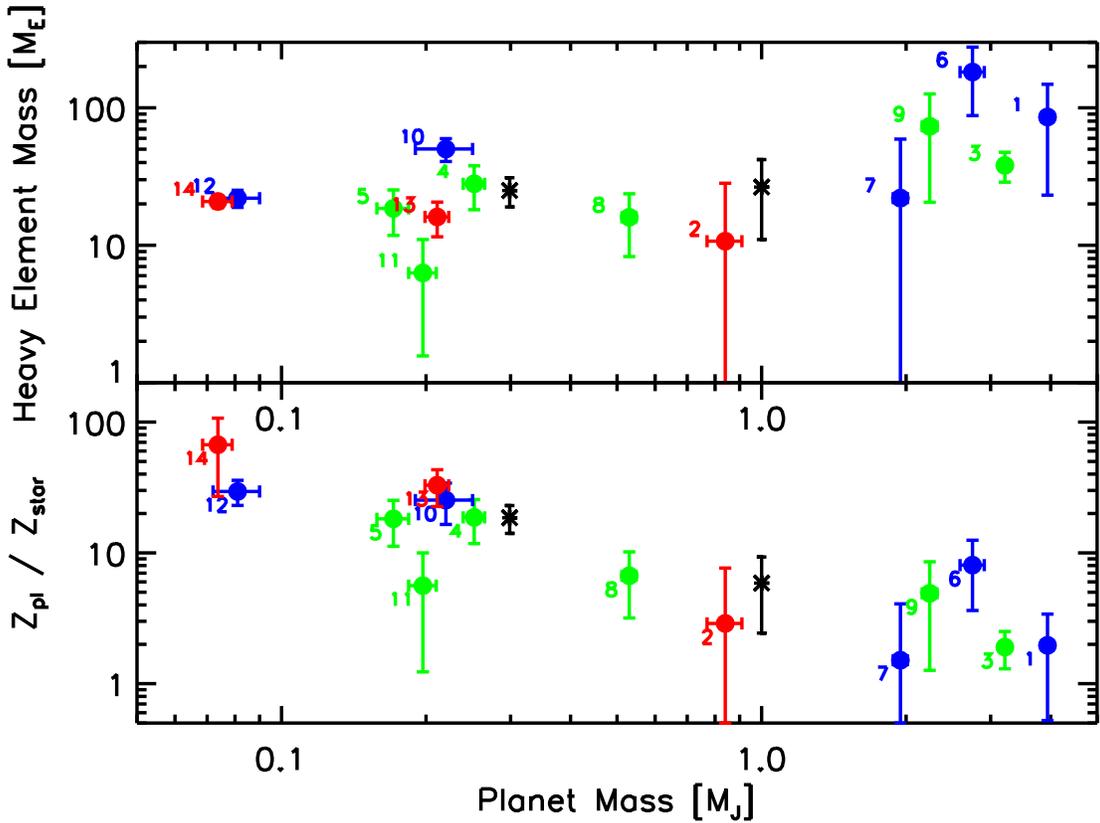}
\caption{
Top: The planet mass and heavy element mass for our sample.  Planets are colored by metallicity in three bins:
$\textrm{[Fe/H] } < 0.0$ (red), 
$0 \le \textrm{ [Fe/H] } < 0.2$ (green), and
$\textrm{[Fe/H] } \ge 0.2$ (blue).  Jupiter and Saturn are also shown in black \citep{Guillot99}.  This plot is consistent with a minimum heavy element mass of 10-15 $M_{\oplus},$
with increasing heavy element masses for larger mass planets.
Bottom: The planet mass and heavy element enrichment ratio $Z_{\textrm{pl}}/Z_{\textrm{star}}$.  
Lower mass planets are more metal enriched, but have less total heavy elements, which is consistent with the solar system's giants \citep{FortneyNettelmann10}.
}
\end{center}
\end{figure}

\begin{sidewaystable*}[!htbp]
\tiny
\caption{Table of Planets with low incident flux}
\centering
\begin{tabular}{r|c c c c c c c c c c c l}
\hline\hline
Number & Name & Mass          & Radius       & Age & $\langle F \rangle$    & [Fe/H] & Core Model & Mixed Model & Average Model & $Z_{\textrm{pl}}$ & $Z_{\textrm{pl}}/Z_{\textrm{star}}$ & References \\
       &      & ($M_{\rm J}$) &($R_{\rm J}$) &Gyr  & erg s$^{-1}$ cm$^{-2}$ &        & ($M_{\oplus}$) & ($M_{\oplus}$) & ($M_{\oplus}$) & & & \\
\hline
\hline
 1 & HD 80606 b &  3.940$\pm$ 0.110 &  1.030$\pm$ 0.036 &   7.0$\pm^{  4.0}_{  4.0}$ & 1.67 $ \times 10^{ 7}$ &    0.390$\pm$   0.030 &  87.0$\pm$ 62.6 &  84.5$\pm$ 62.6 &  85.8$\pm$ 62.6 &    0.068$\pm$   0.050 &      2.0$\pm$     1.4 & [A],[A'] \\
 2 & CoRoT-9 b &  0.840$\pm$ 0.070 &  1.050$\pm$ 0.040 &   4.0$\pm^{  5.0}_{  3.0}$ & 6.58 $ \times 10^{ 6}$ &   -0.010$\pm$   0.060 &  11.1$\pm$ 17.6 &  10.3$\pm$ 17.6 &  10.7$\pm$ 17.6 &    0.040$\pm$   0.066 &      2.9$\pm$     4.8 & [B] \\
 3 & HD 17156 b &  3.212$\pm$ 0.007 &  1.087$\pm$ 0.006 &   3.4$\pm^{  0.6}_{  0.4}$ & 1.96 $ \times 10^{ 8}$ &    0.140$\pm$   0.080 &  38.4$\pm$  9.3 &  37.8$\pm$  9.3 &  38.1$\pm$  9.3 &    0.037$\pm$   0.009 &      1.9$\pm$     0.6 & [C] \\
 4 & Kepler-9 b &  0.252$\pm$ 0.013 &  0.842$\pm$ 0.069 &   3.0$\pm^{  1.0}_{  1.0}$ & 8.11 $ \times 10^{ 7}$ &    0.120$\pm$   0.040 &  31.0$\pm$  9.4 &  25.0$\pm$  9.4 &  28.0$\pm$  9.9 &    0.349$\pm$   0.124 &     18.7$\pm$     6.9 & [D] \\
 5 & Kepler-9 c &  0.171$\pm$ 0.013 &  0.823$\pm$ 0.067 &   3.0$\pm^{  1.0}_{  1.0}$ & 3.14 $ \times 10^{ 7}$ &    0.120$\pm$   0.040 &  20.6$\pm$  6.4 &  16.5$\pm$  6.4 &  18.5$\pm$  6.7 &    0.341$\pm$   0.127 &     18.2$\pm$     7.0 & [D] \\
 6 & CoRoT-10 b &  2.750$\pm$ 0.160 &  0.970$\pm$ 0.070 &   2.0$\pm^{  1.0}_{  1.0}$ & 5.38 $ \times 10^{ 7}$ &    0.260$\pm$   0.070 & 192.0$\pm$ 93.8 & 172.3$\pm$ 93.8 & 182.1$\pm$ 94.4 &    0.208$\pm$   0.109 &      8.1$\pm$     4.4 & [E] \\
 7 & HAT-P-15 b &  1.946$\pm$ 0.066 &  1.072$\pm$ 0.043 &   6.8$\pm^{  2.2}_{  1.8}$ & 1.51 $ \times 10^{ 8}$ &    0.220$\pm$   0.080 &  22.6$\pm$ 37.1 &  21.5$\pm$ 37.1 &  22.0$\pm$ 37.1 &    0.036$\pm$   0.060 &      1.5$\pm$     2.6 & [F] \\
 8 & HAT-P-17 b &  0.530$\pm$ 0.018 &  1.010$\pm$ 0.029 &   7.8$\pm^{  2.2}_{  2.8}$ & 8.91 $ \times 10^{ 7}$ &    0.000$\pm$   0.080 &  16.9$\pm$  7.7 &  15.1$\pm$  7.7 &  16.0$\pm$  7.7 &    0.095$\pm$   0.046 &      6.7$\pm$     3.5 & [G] \\
 9 & WASP-8 b &  2.240$\pm$ 0.080 &  1.038$\pm$ 0.047 &   4.0$\pm^{  1.0}_{  1.0}$ & 1.79 $ \times 10^{ 8}$ &    0.170$\pm$   0.070 &  76.6$\pm$ 52.8 &  70.1$\pm$ 52.8 &  73.4$\pm$ 52.8 &    0.103$\pm$   0.074 &      4.9$\pm$     3.6 & [H] \\
10 & CoRoT-8 b &  0.220$\pm$ 0.030 &  0.570$\pm$ 0.020 &   3.0$\pm^{  1.0}_{  2.0}$ & 1.22 $ \times 10^{ 8}$ &    0.300$\pm$   0.100 &  55.2$\pm$  8.1 &  45.2$\pm$  8.1 &  50.2$\pm$  9.5 &    0.717$\pm$   0.167 &     25.3$\pm$     8.8 & [I] \\
11 & HAT-P-18 b &  0.197$\pm$ 0.013 &  0.995$\pm$ 0.052 &  12.4$\pm^{  4.4}_{  6.4}$ & 1.18 $ \times 10^{ 8}$ &    0.100$\pm$   0.080 &   6.8$\pm$  4.7 &   5.8$\pm$  4.7 &   6.3$\pm$  4.7 &    0.100$\pm$   0.076 &      5.6$\pm$     4.4 & [J] \\
12 & HAT-P-11 b &  0.081$\pm$ 0.009 &  0.422$\pm$ 0.014 &   6.5$\pm^{  5.9}_{  4.1}$ & 1.31 $ \times 10^{ 8}$ &    0.310$\pm$   0.050 &  23.5$\pm$  2.7 &  20.4$\pm$  2.7 &  22.0$\pm$  3.1 &    0.854$\pm$   0.154 &     29.5$\pm$     6.4 & [K] \\
13 & HAT-P-12 b &  0.211$\pm$ 0.012 &  0.959$\pm$ 0.030 &   2.5$\pm^{  2.0}_{  2.0}$ & 1.90 $ \times 10^{ 8}$ &   -0.290$\pm$   0.050 &  17.7$\pm$  4.2 &  14.5$\pm$  4.2 &  16.1$\pm$  4.5 &    0.239$\pm$   0.069 &     32.9$\pm$    10.3 & [L] \\
14 & GJ 436 b &  0.074$\pm$ 0.005 &  0.377$\pm$ 0.009 &   6.0$\pm^{  4.0}_{  5.0}$ & 4.03 $ \times 10^{ 7}$ &   -0.030$\pm$   0.200 &  22.1$\pm$  1.6 &  19.5$\pm$  1.6 &  20.8$\pm$  2.1 &    0.888$\pm$   0.108 &     67.0$\pm$    40.0 & [M] \\
\hline
\end{tabular}
\label{plTbl}

\footnotetext*[1]{References: [A]=\citep{Hida10}, [A']=\citep{Nord08}, [B]=\citep{Deeg10}, [C]=\citep{Nutz10}, [D]=\citep{Hol10}, [E]=\citep{Bon10},
[F]=\citep{Kov10}, [G]=\citep{How10}, [H]=\citep{Que10}, [I]=\citep{Bor10}, [J]=\citep{Hart10}, [K]=\citep{Bak10}, [L]=\citep{Hart09}, [M] = \citep{Torr08}}
\end{sidewaystable*}

\section{Discussion}
The lower-irradiation planets are important for a better understanding of 
giant planet structure and formation since they allow us to probe the composition independent of major assumptions.  
By studying the relationship between stellar metallicity, planet mass, and heavy element mass within a planet, we will be able to test predictions of any planet formation theory, and
specifically against predictions from population synthesis models such as \citet{IdaLin2010, Mordasini2009, Thommes2008}.  
These models are now being extended beyond just planet mass vs.~period, to explore the relationship between [Fe/H] and planet heavy element mass \citep{Mordasini2009}.

The core accretion formation mechanism requires that heavy elements form a core of around $\sim$~10 $M_{\oplus}$, which is followed by the accretion of gas from the protoplanetary disk \citep{Pollack96}.   
Some models for Jupiter indicate that most of its heavy elements are in its envelope, not in its core \citep{FortneyNettelmann10}.  Alternatively, gravitational instability could result in enhanced metallicity through planetesimal accretion \citep{Guillot00, Helled06, Helled2009, Helled10, Boley10, Nayakshin10}.   
The planets in our population, even the more massive planets, are consistent with being enhanced in heavy elements relative to their parent star.  
This enrichment in heavy elements is a distinguishing characteristic between planets and low mass brown dwarfs with more solar-like abundances \citep{Chabrier07b,Leconte2009}.  

If this emerging relationship between stellar metallicity and planetary heavy elements continues to hold with additional data, then
the relationship could be used to determine the amount of heavy elements in a given \emph{inflated} hot Jupiter with some confidence, based only on the parent star metallicity and planet mass.  This would be powerful as it would allow for a straightforward determination of the additional energy needed to explain a planet's inflated radius.  The additional energy source could then be derived for
each inflated planet, as a function of planet mass and irradiation level, and could be compared to
model predictions \citep{Guillot02,Batygin2011}.

As additional data accumulate, modifications to the relations presented here could be in order.   
Perhaps a spread in $Z_{\textrm{pl}}/Z_{\textrm{star}}$ could be due to orbital period, which could tie into the planet's dynamical environment \citep{Guillot00}.  These relationships may be interesting
to analyze in systems with multiple transiting planets.
Another aspect related to orbital evolution is possible:  Perhaps differences in heavy elements could be seen between planets that are well-aligned or mis-aligned with their stellar spin axis, as measured by the Rossiter-McLaughlin Effect \citep{GaudiWinn2007}, as these planets may have taken different paths to their current orbits.  The relationship to stellar mass could also be investigated.

The planet formation process and composition may also be a function of the types of heavy elements
that are in a protoplanetary disk.  
Previously, \citet{Robinson2006} showed empirically that [Si/Fe] or [Ni/Fe] are correlated with the existence of planets for a fixed [Fe/H].  Theoretically, they also showed that ice-rich disks tend to form cores faster.  
Therefore, as the sample of cooler planets in this domain increases, it will also be interesting
to test how these planet composition trends are a function of [$\alpha/H$] or on [Si/H], [O/H] or [C/H].  It may be possible to constrain the composition of the planetary heavy elements from such studies.

In closing, we find evidence from a sample of 14 transiting giant planets that these planets, as a class, are enhanced in heavy elements.
The large heavy element abundances found indicate that all heavy elements cannot be found solely in a core.  If the solar system and planet formation models are a guide, then, in addition to their dense cores, the H/He envelopes of these planets will be enhanced in heavy elements as well, which can be tested by observations of the atmospheres of planets via transit or direct imaging spectroscopy \citep[see, e.g.][]{Marley07b}.

The trends identified here, that independent of stellar metallicity, all giant planets have a heavy element mass of 10 $M_{\oplus}$ or larger, that the abundance of heavy elements in giant planets increases steeply with stellar metallicity, that Jupiter-like enhancement over solar abundances are standard for gas giants, and that more massive planets tend to have lower enrichment, could be enhanced or refuted by additional detections of transiting planets with equilibrium temperatures less than 1000 K.  These longer-period systems will continue to be detected from the ground and recently NASA's \emph{Kepler} spacecraft identified dozens of candidates for a potentially dramatically larger sample of these less-irradiated transiting giant planets \citep{Borucki11}.

\acknowledgements
We thank Mark Marley, Kevin Schlaufman, James Guillochon, Philip Nutzman, and Eliza Kempton for
providing feedback and encouragement.  JJF acknowledges the support of NSF grant AST-1010017 and an Alfred P.~Sloan Research Fellowship.


\end{document}